# An Overview and Comparison of Axiomatization Structures Regarding Inconsistency Indices' Properties in Pairwise Comparisons Methods

Sangeeta Pant [a], Anuj Kumar [b], Jiří Mazurek [c*]

[a] Department of Applied Sciences, Symbiosis Institute of Technology, Symbiosis International (Deemed University) (SIU), Lavale, Pune 412115, Maharashtra, India. E-mail: pant.sangeet@gmail.com

[b] School of Computer Science Engg. & Applications, D. Y. Patil International University (DYPIU), Akrudi, Pune 411044, Maharashtra, India. E-mail: anuj4march@gmail.com

[c] Department of Informatics and Mathematics, School of Business Administration in Karvina, Silesian University in Opava. E-mail: mazurek@opf.slu.cz

[*] corresponding author

**Abstract:** Mathematical analysis of the analytic hierarchy process (AHP) led to the development of a mathematical function, usually called the *inconsistency index*, which has the center role in measuring the inconsistency of the judgements in AHP. Inconsistency index is a mathematical function which maps every pairwise comparison matrix (PCM) into a real number. An inconsistency index can be considered more trustworthy when it satisfies a set of suitable properties. Therefore, the research community has been trying to postulate a set of desirable rules (axioms, properties) for inconsistency indices. Subsequently, many axiomatic frameworks for these functions have been suggested independently, however, the literature on the topic is fragmented and missing a broader framework. Therefore, the objective of this article is twofold. Firstly, we provide a comprehensive review of the advancements in the axiomatization of inconsistency indices' properties during the last decade. Secondly, we provide a comparison and discussion of the aforementioned axiomatic structures along with directions of the future research.

## 1. Introduction

Pairwise comparisons are cornerstones and the most essential part of many multicriteria decision making methods such as the AHP (*Analytic Hierarchy Process*) (Saaty, 1977, 1980, 2004), BWM (*Best-Worst Method*) (Rezaei, 2015), ELECTRE (*ÉLimination Et Choix Traduisant la REalité i.e., Elimination and Choice Translating Reality*) (Roy, 1968), MACBETH (*Measuring Attractiveness by a Categorical Based Evaluation* Technique) (Bana e Costa & Vansnick, 1994;

Bana e Costa et al., 2012), or PROMETHEE (*Preference Ranking Organization Method for Enrichment of Evaluations*) (Vincke & Brans, 1985). Psychometricians had been performing pairwise comparisons for a very long period prior to the discovery of AHP, see e.g. the study of L. L. Thurstone entitled *A Law of Comparative Judgement* from 1927 (Thurstone, 2017). According to psychologists, it is more convenient and easier to make some solid opinions (comparisons) about only two alternatives than simultaneously about all the given alternatives. This led to the development of the theory of pairwise comparison called *paired comparison* in the field of psychology. These pairwise comparisons are mathematically represented by a positive pairwise comparison matrix (PCM). The ability of pairwise comparisons methods to break down large decision-making problems into manageable sub-problems is their fundamental strength. The decision-maker (DM) is empowered by these pairwise comparisons to recognize the flaws in his decisions, which enables DM to be more rational when making the final judgements. The consistency of the judgements can be mathematically identified by a real valued function named as inconsistency index.

There are two types of inconsistencies: *ordinal inconsistency* (weaker consistency condition) and *cardinal inconsistency* (stronger consistency condition). In ordinal pairwise comparisons, the only information that is provided is the relative preference between two objects without any numerical values. This means that if an object A is preferred to an object B, and the object B is preferred to an object C, then the object A should be preferred to the object C. This *transitive property* is essential for ensuring that the preferences are logically consistent and do not contain any contradictions. Moreover, if transitivity of preferences is violated, then it is not possible to (partially or totally) rank compared objects. Ordinal pairwise comparisons are often used in decision-making processes where numerical values cannot be assigned to objects, such as when evaluating subjective preferences or qualitative characteristics of objects.

Ordinal inconsistency implies cardinal inconsistency; however, the converse is not true. It is possible to have cardinal inconsistency without ordinal inconsistency if the numerical values assigned to the objects do not reflect the underlying ordinal preferences accurately (Yang et al., 2016; Siraj et al., 2012). This article focuses solely on the developments of the axiomatic structures for cardinal inconsistency.

Consistency issues in AHP are pivotal in the debate of the mathematical analysis of AHP. Saaty (1980) introduced two mathematical functions known as *consistency index* ($CI$) and *consistency ratio* ($CR$), which have been widely used by the decision makers to evaluate the inconsistency in the AHP.

Criticism on the threshold of the $CR$ index (Kuenz, 1993; Lane et al., 1989; Salo & Hämäläinen, 1997); Ágoston & Csató, 2022) has given a fertile ground to develop new inconsistency indices. Since then, tens of mathematical functions to evaluate inconsistency of judgements in PC methods have been developed independently and empirically (Brunelli, 2018; Pant et al. 2022), and researchers are still developing new indices to measure the inconsistency in PCM.

However, there is still a room for discussion regarding the applicability of already created inconsistency indices. Perhaps, Saaty's seminal work "A scaling method for priorities in hierarchical structures" (Saaty, 1977) can be regarded as independent from other works on inconsistency indices. In literature, some of the inconsistency indices have been introduced which are similar or mathematically related to each other (Brunelli et al., 2013; Cavallo, 2017; Cavallo, 2020; Fedrizzi & Ferrari, 2018; Brunelli et al., 2013). Therefore, some of their properties are similar or even identical with each other. Saaty's inconsistency ratio and Koczkodaj's inconsistency index on the set of triads are shown to have a differentiable one-to-one correspondence by Bozóki & Rapcsák (2008). A study by Brunelli (2016) demonstrates the functional relationship between two inconsistency indices that were introduced in two distinct frameworks, making them nearly equivalent. A general framework for establishing inconsistency indices was developed by Bortot et al. (2023), with a parametric generating function being as the system's main building block of overall framework.

When taking into account the inconsistency indices, which are distinct from one another, the axiomatic structure takes on more significance. Axiomatic structure establishes guidelines for a function to be honored as an inconsistency index. The pairwise comparisons matrices have been the subject of numerous studies, most of which concentrate on the measurement scales, inconsistency problems, and priority derivation techniques in multiplicative, additive, or interval (fuzzy) PCM.

The first thing that springs to mind whenever we discuss a specific mathematical concept, such as vector spaces, metric spaces, topological spaces, etc., is that, a mathematical structure must satisfy certain properties to belong to a particular class of spaces. Likewise, every function for assessing the inconsistency of PCM must satisfy a certain set of criteria. Axiomatic analysis has a long tradition in social choice theory since at least Arrow's theorem (Arrow, 1950 and 2012). Similarly, axiomatic discussion is a well-known and fruitful approach for investigating weighting methods of pairwise comparison matrices (Barzilai, 1997; Barzilai, et. al., 1987; Csató, 2018 and 2019; Csató & Petróczy, 2021; Fichtner, 1984 and 1986).

The appropriateness of a function as an inconsistency index may be questioned if these principles or axioms are violated. Because of this, the research community has also moved attention to the need to create a set of axioms that will unify inconsistency evaluation framework.

This paper's aim is to provide an overview and comparison of several axiomatic structures on inconsistency indices' properties in the multiplicative pairwise comparisons framework proposed in the literature over the last decade, and to provide future research directions in this area.

The paper is organized as follows: Section 2 gives a conceptual overview of multiplicative pairwise comparisons and a brief introduction to the inconsistency indices. The various axiomatic structures put forth so far are presented in Section 3 followed by a short graphical summary in section 4. The study's conclusion and recommendations for the future research are provided in the last Section 5.

## 2. Multiplicative Pairwise Comparisons

In this section notation and important concepts of multiplicative pairwise comparisons are introduced.

Let $n \in \mathbb{N}, n \geq 2$ be the set of compared objects (usually alternatives, criteria, sub-criteria, etc.). Let $S$ be a multiplicative (ratio) scale. Then every $a_{ij} \in S$ expresses how many times is an object *i* more preferred (important) over an object *j*. In particular, $a_{ij} > 1$ means the object *i* is preferred over the object *j*, and $a_{ij} = 1$ means both objects are of the same preference.

All $n \times n$ pairwise comparisons can be arranged into a square pairwise comparisons (PC) matrix $A = [a_{ij}]$ of the order *n*:

$$A = \begin{bmatrix} a_{11} & a_{12} & \cdots & a_{1n} \\ a_{21} & a_{22} & \cdots & a_{2n} \\ \vdots & \vdots & \ddots & \vdots \\ a_{n1} & a_{n2} & \cdots & a_{nn} \end{bmatrix}$$

**Definition 1**. Let $A = [a_{ij}]$ be a PC matrix of the order n. Then the matrix A is reciprocal if:

$$a_{ij} = \frac{1}{a_{ji}}, ; \forall \, i, j. \tag{1}$$

It is commonly assumed that a PC matrix is reciprocal.

**Definition 2**. Let $A = [a_{ij}]$ be a PC matrix of the order n. Then the matrix A is cardinally consistent if:

$$a_{ij} \cdot a_{jk} = a_{ik}; \, \forall \, i, j, k. \tag{2}$$

Hence, cardinal inconsistency occurs if for some $i, j, k \in \{1,2,3,\ldots,n\}$, the inequality $a_{ik} \neq a_{ij} \cdot a_{jk}$ holds.

**Definition 3**. The triplet $(a_{ij}, a_{jk}, a_{ik})$ from Definition 2 is called a 'triad'.

**Proposition 1** (Saaty, 1994): For a consistent PC matrix $A = [a_{ij}]$ of the order $n$ the following statements are equivalent:

1. The matrix A is cardinally consistent.
2. There exists a weight vector $w = \{w_1, w_2, w_3, \ldots, w_n\}$ such that $a_{ij} = \frac{w_i}{w_j}$ ; $\forall, i, j$.
3. The spectrum of A consists just of two distinct (real) elements: $n$ and $0$. The algebraic multiplicity of $n$ and $0$ are 1 and $(n-1)$ respectively.
4. The rank of A is one.

Proposition 1 in its property number 2 allows a decision maker to find a priority vector (a vector of weights of all compared objects) so that the best object can be found, or, alternatively, all objects can be ranked from the best to the worst. The property number 4 in Proposition 1 can provide a fast check of matrix consistency. The property number 3 leads directly to the definition of Saaty's consistency index *CI*:

**Definition 4** (Saaty, 1980). Let $A = [a_{ij}]$ be a PC matrix of the order n. Then the consistency index CI is given as follows:

$$CI = \frac{\lambda_{max} - n}{n-1}, \tag{3}$$

where $\lambda_{max}$ is the largest (positive and real) eigenvalue of the matrix *A*. When a PC matrix is consistent, then $\lambda_{max} = n$, hence *CI* = 0. Otherwise, $\lambda_{max} > n$, hence *CI* > 0.

From the *CI* index, Saaty's consistency ratio *CR* can be readily obtained as $CR = \frac{CI}{RI}$, where *RI* denotes average value of *CI* for randomly generated matrices of the given order *n*. Per the AHP nomenclature, PC matrices with *CR* < 0.10 are tolerated.

Other inconsistency indices can be found in Mazurek (2023).

**Definition 5**. *Let $C = \{c_1, \ldots, c_n\}$ be a set of concepts. Let '≻' denote the relation of preference. Then $c_i ≻ c_j$ expresses that the concept $c_i$ is preferred over an object $c_j$. Preferences are transitive if the following condition holds: $c_i ≻ c_j \wedge c_j ≻ c_k \Rightarrow c_i ≻ c_k$.*

By applying Definition 5 on PC matrices we get the following corollary.

**Corollary 1**. *Let $A = [a_{ij}]$ be a PC matrix of the order n. Then the triad $(a_{ij}, a_{jk}, a_{ik})$ is transitive if the following condition holds: $a_{ij} > 1 \wedge a_{jk} > 1 \Rightarrow a_{ik} > 1$.*

**Definition 6**. *Let $A = [a_{ij}]$ be a PC matrix of the order n. Then the matrix A is ordinally consistent if all triads contained in the matrix A are transitive.*

Pairwise comparisons acquired from human experts are often inconsistent, see e.g. (Linares et al., 2016; Mazurek & Perzina, 2017; Mazurek, & Neničková, 2020). Therefore, the need of measuring of the extent of inconsistence emerged.

Cardinal inconsistency is usually measured via a suitable *inconsistency index*.

**Definition 7**. *Let $A = [a_{ij}]$ be a PC matrix of the order n and let $A_n$ be the set of all pairwise comparison matrices of the order n. The, an inconsistency index I is a function from $A_n$ to $\mathbb{R}$.*

The first inconsistency index was proposed by Kendall & Smith, (1940). Since then, several inconsistency indices have been suggested in the literature such as Saaty's eigenvalue-based index *CI* (Saaty, 1988), the Koczkodaj inconsistency index (Koczkodaj, 1993; Duszak & Koczkodaj, 1994), the geometric consistency index (Crawford, 1987; Aguarón & Moreno-Jiménez, 2003), Cosine consistency index (Kou & Lin, 2014; Khatwani & Kar, 2017), Salo-Hämäläinen index (Salo & Hämäläinen, 1997), the Lamata and Peláez index (Lamata & Peláez, 2002; Peláez & Lamata, 2003), or Harmonic consistency index (Stein & Mizzi, 2007). There is a direct impact on the *CI* index if we change a matrix element, as proved by Aupetit & Genest, (1993). The *CI* must either be always rising, always decreasing, or dropping to a minimum before increasing if any upper triangular entry of the matrix increases. As a result, the *CI* function should have a unique local minimum. The preceding judgments must be revised if the consistency measure is higher than the threshold value but at the same time, there are many applications of pairwise comparisons as well where the judgements cannot be reconsidered because they are derived from exogenous objective data (Bozóki et al., 2016; Chao et al., 2018; Csató, 2013; Csató & Tóth, 2020; Petróczy, 2021; Temesi et al., 2023). A meaningful interpretation is difficult to offer if there is no threshold linked with the existing consistency measurement. Without a consistency threshold, a decision maker is left with the difficult decision of when to revise and when to accept their judgments.

Many inconsistency indices, however, have been suggested in the literature without mentioning the thresholds connected to them. Additionally, different indices frequently produce contrasting findings. Using numerical simulation, Brunelli (2016) has shown the discordant behavior of the two indices proposed by Barzilai (1998) and Gass & Rapcsák (2004), respectively. To demonstrate the inconsistency indices' reliability as an estimator of inconsistency, all these concerns necessitate their proper axiomatization.

### 3. Axiomatization of consistency indices

Koczkodaj & Szwarc (2014) established three axioms for the local consistency of triads in a multiplicative PCM. They proposed the term "triad" to refer to any three elements from the abelian group $\mathbb{R}^+$ (under the usual multiplication operations) of positive real numbers $a, b$ and $c$. A square matrix $A = [a_{ij}]$ of order $n$ is called a matrix over the group $\mathbb{R}^+$ if every element of the matrix belongs to the set $\mathbb{R}^+$. Hence, every multiplicative PCM is a matrix over the group $\mathbb{R}^+$.

**Definition 8** *(Koczkodaj & Szwarc, 2014). Let $(a, b, c)$ denote a triad from the space $\mathbb{R}^{+3}$, then according to a function $f : \mathbb{R}^{+3} \to \mathbb{R}$ can be regarded as inconsistency index if it meets the following three axioms:*

-Axiom 1: *$f$ maps a consistent triad $(a, b, c) \in \mathbb{R}^{+3}$ to 0, i.e., $f(a, b, c) = 0$.*

-Axiom 2: *There exists no inconsistency index which maps a triad $(a, b, c)$ to 1, i.e., complete inconsistency can never be achieved, thus the codomain of the function $f$ is $[0,1[$.*

-Axiom 3: *If $(a, b, c)$ is a consistent triad i.e. $f(a, b, c) = 0$ and if the values of $a, b$ or $c$ are altered, then the new triad $(a', b', c')$ may no longer remain consistent, i.e., $f(a, b, c) > 0$. (Thus, a consistent triad becomes necessarily inconsistent when any of its values is altered).*

The axioms from Definition 8 are specified for the local character of an inconsistency index. The first axiom results from the consistency requirement stated by Kendall & Smith (1940). Koczkodaj (1993) put forth the second Axiom for the first time in 1993. Further on, Koczkodaj et al. (2017) gave a number of reasons to choose the interval [0,1] instead of [0, ∞[ as the range set of inconsistency indices. Naturally, normalization makes it easier to use the consistency index. Further on, Koczkodaj et al. (2020) argued that only real numbers should be used as a range of an inconsistency index.

This axiomatic structure was distinctive because no previous investigation on inconsistency indices had been conducted, but Brunelli (2016) raised some concerns about some of the shortcomings of these axioms. The negative aspects were as follows:

1. Axioms were primarily designed to be applied to a single triad, making them appropriate for a matrix of order $3 \times 3$. The PCM of an order larger than 3 was not explicitly covered by these axioms.
2. The third axiom did not take into account all instances of increasing/decreasing of values for $a$ and $c$. Just two cases—where either $a$ or $c$ are increasing or decreasing—were used by Koczkodaj & Szwarc (2014). The latter two scenarios — where $a$ increases while $c$ decreases and $a$ decreases while $c$ increases — were not investigated.
3. The three axioms' independence from one another was not established.

"It is a reasonable expectation that the worsening of a triad, used in the definition of consistency, cannot make the entire matrix more consistent", Koczkodaj & Szwarc (2014) said in support of the third axiom.

Brunelli (2016) disputed this assumption and provided a refuting example. In that example, when one triad's consistency was worsened, the consistency of the other two triads was improved, which in turn also enhanced the matrix's consistency. A pairwise comparison matrix of order $n \times n$ actually has any entry $a_{ij}$ appear in $(n-2)$ triads of the matrix. The consistency of the remaining $(n-1)$ triads won't always degrade if the consistency of one of the triads containing $a_{ij}$ declines. In fact, Brunelli (2016) demonstrated in a counter example of a matrix of order $5 \times 5$ that when the consistency of one triad $(a_{12}, a_{15}, a_{25})$ containing $a_{15}$ was worsened, the consistency of all other triads $(a_{13}, a_{15}, a_{35})$ and $(a_{14}, a_{15}, a_{45})$ containing $a_{15}$ was strengthened.

There is no doubt that a set of positive reciprocal matrices does not constitute a vector space over the field $\mathbb{R}$ of real numbers. In this article, $A_n(\mathbb{R}^+)$ denotes the set of all positive reciprocal matrices of order $n$, which is defined as follows.

Thus, an inconsistency index $f$ is a function from the set $A_n(\mathbb{R}^+)$ to the set of real numbers $\mathbb{R}$. Brunelli (2016) re-structured the aforementioned axioms specifically for the pairwise comparison matrices $[a_{ij}] = A \in A_3(\mathbb{R}^+)$ as follows:

***Axiom 1:*** *For a consistent matrix $A \in A_3(\mathbb{R}^+)$, $f(A) = 0$.*

***Axiom 2:*** *For an inconsistent matrix $A \in A_3(\mathbb{R}^+)$, $f(A) \in ]0,1[$.*

***Axiom 3:*** *The function $f$ is a quasi-convex function of three variables $a_{12}, a_{13}$ and $a_{23}$ and its absolute minimum occurs whenever $a_{12}. a_{23} = a_{13}$.*

The next step in the construction of axioms was made by Brunelli & Fedrizzi (2015). Five sound axioms were introduced as part of their axiomatic structure. Nevertheless, its initial publication, Brunelli & Fedrizzi (2011), contained the first four attributes. Instead of focusing on the inconsistency of a triad, these axioms were more concerned with the inconsistency of a matrix of any order. The following succinct summary of these five axioms proposed by Brunelli & Fedrizzi (2015) is as follows:

***Axiom 1:*** *Inconsistency index $f$ must map all consistent PCM to a unique real number $\omega$. That is, A is consistent if and only if $f(A) = \omega$.*

Two distinct consistent matrices must be mapped to the same real number $\omega$ by an inconsistency index. Inconsistent matrices, on the other hand, must be mapped to set $\mathbb{R}\backslash\{\omega\}$.

This axiom's mathematical meaning is similar to that of first axiom in (Koczkodaj & Szwarc, 2014). While in Koczkodaj & Szwarc (2014) the minimum normalized value of the inconsistency index "0" was taken as a representative of the consistent matrix, Brunelli & Fedrizzi (2015) considered the minimum value of the function $f$ to be the value corresponding to the consistent matrix.

***Axiom 2:*** *The PCM's inconsistency assessment must be unaffected by the order in which the alternatives are assigned to its rows and columns. i.e.,*

$$(\forall A \in A_n(\mathbb{R}^+)) f(P^T A P) = f(A),$$

*where P is a permutation matrix.*

The aforementioned axiom has the mathematical meaning that $A$ and $P^T A P$ must have the same value of inconsistency because they are similar matrices and as a result, must have the same eigenvalues. Depending on the structure of a permutation matrix $P$, pre- and post-multiplications of $P$ with A will result in the rows and columns being switched, respectively. Hence, the order of alternatives will vary as a result.

***Axiom 3:*** *Let $A = [a_{ij}]$ is a PC matrix and for a given $k > 1$, $A' = [a_{ij}^k]$, then the inconsistency measure of $A'$ must be greater than that of $A$.*

It is obvious that shifting preferences in a PCM away from the neutral value, i.e., $a_{ij} = 1, \forall i, j$, leads to a higher value of inconsistency. This axiom can be interpreted simply as saying that as preferences become more intense, inconsistency will increase in severity. According to Brunelli & Fedrizzi (2015), $g(x) = x^k, k > 1, x > 0$ is the only continuous monotonic function that intensifies the preferences while preserving both their structure and the consistency of the PCM. This function serves to intensify the preferences since it maps every non-unity PCM entry farther away from indifference value 1. Therefore, the new matrix $A'$ must be more inconsistent than the old matrix $A$, i.e., $f(A') \geq f(A)$. Equality holds only when the matrix $A$ is consistent.

***Axiom 4:*** *An inconsistency index f satisfies Axiom 4 iff:*

   a. *If $a_{ij}'' > a_{ij}' > a_{ij}$ then $f(A'') \geq f(A') \geq f(A)$.*
   b. *If $a_{ij}'' < a_{ij}' < a_{ij}$ then $f(A'') \geq f(A') \geq f(A)$.*

*where, $a_{ij}$ is a non-diagonal element of some consistent matrix A, and $A'$ and $A''$ are the two matrices that are generated by altering the value of $a_{ij}$ to $a_{ij}'$ and $a_{ij}''$, respectively.*

Therefore, altering a single non-diagonal element $a_{ij}$ of a consistent PCM "$A$" leads to an inconsistent matrix with non-decreasing inconsistency the further away from $a_{ij}$ we proceed. This axiom follows from the conclusion proved by Aupetit & Genest (1993): let $a_{ij}$ be a fixed nondiagonal entry of a consistent matrix $A$ and if $\Delta a_{ij} = a_{ij} - a_{ij}^{new}$ then $f$ is a monotonically increasing function of $\Delta a_{ij}$ when $\Delta a_{ij} > 0$ and $f$ is a monotonically decreasing function of $\Delta a_{ij}$ when $\Delta a_{ij} < 0$.

***Axiom 5:*** *An inconsistency function f meets axiom 5 if and only if the function f is a continuous function of $\frac{n^2-n}{2}$ real variables, i.e., a function of the upper half entries $a_{ij}, i < j$ of a PCM.*

The primary distinction between the axiomatizations offered by Koczkodaj & Szwarc (2014) and Brunelli & Fedrizzi (2015) is that former one did not take the continuity of the inconsistency index into account. The ability to grasp how variations in output (inconsistency) correspond to minute changes in input values (preferences) makes continuity an essential component. Barzilai (1998) has made reference to the significance of continuity of the inconsistency index, stating that "continuity of a consistency index $f$ is a legitimate prerequisite of any measure as it is difficult to comprehend discontinuities in this context". Pairwise comparisons are the mathematical representation of psychophysical theory and continuity of psychophysical function was also assumed in the literature (Roberts, 1985). Since there exist discontinuous functions in literature that serve as inconsistency indices, the question of the continuity of inconsistency indices is still up for debate.

The inconsistency indices suggested by Saaty (1980), Aguarón & Moreno-Jiménez (2003), and Lamata & Peláez (2002) meet these 5 axioms. While not all axioms were satisfied by the inconsistency indices suggested in Stein & Mizzi (2007), Barzilai (1998), Golden & Wang (1989) and Ramík & Korviny (2010). Although the consistency index proposed in Barzilai (1998) was considered a continuous function, Brunelli & Fedrizzi (2015) proved that it was actually a discontinuous function.

Mazurek (2017) added one more axiom in the axiomatic structure of the five axioms presented by Brunelli & Fedrizzi (2015). An upper bound of inconsistency indices were the emphasis of this newly added sixth axiom. According to Brunelli & Fedrizzi (2015), a consistent matrix takes the minimum value of an inconsistency index. The lower bound of any inconsistency index is thus always known. The issue is the upper bound of the inconsistency index. According to this new axiom:

"For a set $X$, any function $f: X \to \mathbb{R}$ is called bounded from above if there exists $K \in \mathbb{R}$ such that $(\forall x \in \mathbb{R}) f(x) \leq K$. In particular, an inconsistency index $f: A_n(\mathbb{R}^+) \to \mathbb{R}$, is called bounded from above if there exist $K \in \mathbb{R}$ such that $(\forall A \in A_n(\mathbb{R}^+)) f(A) \leq K$."

Mazurek (2017) then used a corner pairwise comparison matrix to show that the consistency index ($CI$) proposed by Saaty (1980), the geometric consistency index ($GCI$) by Aguarón & Moreno-Jiménez (2003), and the inconsistency index by Lamata & Peláez (2002) ($CI^*$) are not bounded from above.

One more axiom was later added by Brunelli & Fedrizzi (2019) to their previous axiomatic structure:

***Axiom 6:*** *An inconsistency index $f$ satisfies axiom 6 if and only if $f(A) = f(A^T)$, i.e., the consistency ought to remain unchanged when preferences are inverted in a matrix A.*

The aforementioned six axioms, which were introduced by Brunelli & Fedrizzi (2015) and Brunelli (2019), were shown to be independent from one another and non-contradictory. These six axioms are reliable and help to better define the characteristics of an inconsistency index.

Further on, the requirement established by Brunelli & Fedrizzi (2019) was crucial for the particular class of inconsistency indices, such as the $GCI$, $KI$, $CI^*$, $I_{cd}$ and $ATI$ (Grzybowski, 2016), which all share a common structure, i.e., they calculate the global inconsistency by aggregating the local inconsistencies. Such inconsistency indices $f$ that can be rewritten as:

$$fA = \oplus_{1 \leq i < j < k \leq n} (F(\eta_{ijk})), \eta_{ijk} = a_{ij} a_{jk} a_{ki},$$

where $F: \mathbb{R}^+ \to \mathbb{R}$ measures the local inconsistencies (inconsistency of triads of pairwise comparison matrix $A$) and $\oplus: \mathbb{R}^{\binom{n}{3}} \to \mathbb{R}$ is the aggregation function which aggregates these local inconsistencies. Brunelli & Fedrizzi (2019) proved following important result for such a type of inconsistency indices:

**Proposition 2** *(Brunelli & Fedrizzi, 2019). Let F be a function such that:*

1. *F is a continuous, quasi convex in $(0, \infty)$ with the property $(\forall x \in (0, \infty)) F(x) = F\left(\frac{1}{x}\right)$ and the minimum value of F is at $\eta_{ijk} = 1$.*
2. *$\oplus$ is symmetric, monotonically increasing continuous function.*

*then F satisfies all the six axioms given by* Brunelli & Fedrizzi (2015, 2019).

Brunelli (2016b) studied in his conjoint analysis properties both ordinal (in the form of intransitivity) and cardinal inconsistency. Let $A = [a_{ij}]$ be a PCM and suppose $C(A)$ denotes the number of intransitive triplets $(a_{ij}, a_{jk}, a_{ik})$ contained in upper triangle of $A$. Brunelli (2016b) added seventh axiom to his previous system involving intransitivity as follows:

***Axiom 7:*** *Let I be an inconsistency function and let A and B be two PC matrices of the same order n. Then the function I satisfies Axiom 7 if $C(A) \geq C(B)$ implies $I(A) \geq I(B)$.*

Further on, it is shown in Brunelli (2016b) that if a function $I$ satisfies Axiom 1 and Axiom 5, then it cannot satisfy Axiom 7. Also, a non-existence of a function $I$ satisfying all seven axioms is proved. Finally, the author shows that Axioms 1, 2, 3, 4, 6 and 7 are independent and form a logically consistent system.

An enhanced version of Koczkodaj & Szwarc (2014) axiomatic system was presented by Koczkodaj & Urban (2018), along with an additional axiom. Assume $T$ is the collection of all PCMs of finite orders. The following four axioms must be met by a function $f: T \to \mathbb{R}$ in order for it to be the measure of inconsistency of a PCM:

***Axiom 1:*** *A matrix $A \in T$ is consistent if and only if $f(A) = 0$. i.e., the image of all consistent matrices under the map f must have the same value 0.*

The first axiom in the axiomatic system provided by Brunelli & Fedrizzi (2015) and this axiom share a similar mathematical meaning. In Brunelli & Fedrizzi (2015), a single real value $\omega$ must represent the image of every consistent matrix, whereas in this axiom, this value of $\omega$ is set to zero because the normalized range of the inconsistency index was preferred.

***Axiom 2:*** *For an inconsistent matrix A, $f(A) \in ]0,1]$.*

As the normalized range [0,1] of inconsistency function $f$ was adopted, therefore, all inconsistent matrices must be mapped into ]0,1].

***Axiom 3:*** *For every pair of triads $(a_1, b_1, c_1)$ and $(a_2, b_2, c_2)$ inconsistency increases progressively as we move away from the transitivity requirement $b = ac$, i.e.,*

$$|a_1 . c_1 - b_1| \leq |a_2 . c_2 - b_2| \to f(a_1, b_1, c_1) \leq f(a_2, b_2, c_2)$$

*The "exponentially invertible measure" (EIM), which must be topologically equivalent to the distance $(x, y) \mapsto |x - y|$, is a more general measure to assess the deviation from the ideal transitivity condition.*

***Axiom 4:*** *The inconsistency of a PCM can never be less than the inconsistency of its submatrix, i.e., when the collection of compared entities is expanded, we should anticipate an increase in the PC matrix's inconsistency.*

This axiomatic structure took the range set of the inconsistency index [0,1] rather than [0,1[, and did not take into account the ideal inconsistency condition, unlike its earlier version in Koczkodaj & Szwarc (2014). An exponential invertible measure $\Delta_0(a,b) = max\left\{\frac{x}{y}, \frac{y}{x}\right\}$, which is topologically equivalent to the usual distance $d(x,y) = |x - y|$ was used to evaluate the deviation from transitivity condition.

The approach of Csató (2018, 2019) is different from the previous axiomatizations discussed above. Csató (2018, 2019) are, according to our best knowledge, the only studies where a set of properties is proved to uniquely determined an inconsistency index. In particular, Csató (2019) verifies that there exists only one reasonable inconsistency index on the set of triads. All the previous axiomatizations have allowed using more, potentially contradictory, inconsistency indices. This seems to a crucial difference. A PCM is consistent if and only if each of its triads is consistent.

Csató (2019) presented six independent properties of inconsistency indices to axiomatically characterized Koczkodaj inconsistency index. Instead of an $n \times n$ matrix, these features were particularly suggested for triads (except for the Axioms 5 and 6, which are formulated for an arbitrary $n$). Some of these characteristics mirrored those that other authors had already introduced. The following are the six characteristics (axioms):

I. Similar to the fourth axiom of Brunelli & Fedrizzi (2015) of monotonicity on single comparisons, this first axiom of Csató (2019) was stated as follows: Let $(1, a, 1)$ and $(1, b, 1)$ be two inconsistent triads, i.e., $a, b > 1$, and if $a \leq b$ then inconsistency of $(1, a, 1)$ cannot be greater than that of $(1, b, 1)$, i.e.,
$$a \leq b \leftrightarrow f(1, a, 1) \leq f(1, b, 1)$$
That is, the value of inconsistency rises as we get further away from the consistent trio $(1,1,1)$. This notion is referred to as "positive responsiveness".

II. This second axiom is identical to the Axiom 6 ("invariance under inversion of preferences") of Brunelli & Fedrizzi (2019). This second axiom states that a triad's transpose and the original triad both have the same measure of inconsistency.

III. An inconsistency index maps the triads $(1, a, b)$ and $(1, \frac{a}{b}, 1)$ to the same value. If the first and second alternatives are equally significant on their own but are also compared to a third alternative, the inconsistency of the resultant triad should not be affected by the relative importance of the new alternative. This principle is designated as "homogeneous treatment of entities".

IV. Inconsistency index of the triads is independent from the representation of the scale, *i.e.*, $f(a, b, c) = f(ka, k^2b, kc)$

V. This axiom is comparable to the fourth axiom stated by Koczkodaj & Urban (2018). A PCM's triad's (*i.e.*, a submatrix of order 3) inconsistency cannot be bigger than that of the PCM itself. Monotony is the name given to it.

VI. There must be at least one triad whose inconsistency contributes to the PCM's overall inconsistency, *i.e.,* at least one triad whose inconsistency is similar to the PCM's inconsistency. This property was named 'reducibility'.

4. **Summary**

In the previous section, four main axiomatic systems were introduced (see below). Therefore, a natural question arises: how similar are these systems?

Similarity of two sets can be evaluated via Jaccard index (Jaccard, 1912).

**Definition 9**. *Let C and D be two (finite, non-empty) sets. Then the Jaccard index is given as follows:*

$$J(C,D) = \frac{|C \cap D|}{|C \cup D|} \qquad (4)$$

It is apparent from Definition 9 that $0 \leq J(C,D) \leq 1$. In particular, $J(C,D) = 0$ if the two sets have no intersection, while $J(C,D) = 1$ if the two sets are identical.

The four systems to be evaluated (in their original form) are as follows:

- Koczkodaj & Szwarc (2014) (*KS*): 3 axioms.

- Brunelli & Fedrizzi (2015) (*BF*): 5 axioms.

- Koczkodaj & Urban (2018) (*KU*): 4 axioms.

- Csató (2019) (*CS*): 6 axioms.

The intersection in the numerator of (4) consists of axioms which are identical or almost identical in two given sets. The union in the denominator of (4) is just the sum of the number of axioms in both sets.

The similarity of these systems with respect to Jaccard index is shown in Figure 1. The most similar systems are *KU* and *BF*, and the least similar are *BF* and *CS*.

**Fig. 1**. Similarity of axiomatic systems with respect to Jaccard index. Source: authors.

|    | KS     | KU     | BF     | CS     |
|----|--------|--------|--------|--------|
| KS | 1      | 0.1429 | 0.25   | 0.1111 |
| KU | 0.1429 | 1      | 0.3333 | 0.2    |
| BF | 0.25   | 0.3333 | 1      | 0.0909 |
| CS | 0.1111 | 0.2    | 0.0909 | 1      |

A timeline of advances in the axiomatization of inconsistency indices, in a simplified form, is shown in Figure 2.

Since the list of inconsistency indices is rather long and growing every year, Table 1 provides a summary of inconsistency indices and their properties which have been reported in the literature so far. Here, in Table 1, $A_i$ denotes the $i^{th}$ axiom in the author's suggested set of axioms.

It should be noted that in the most comprehensive publication on the topic, Mazurek (2023), more than 25 inconsistency indices are provided which means that properties of several indices have not been investigated yet. Also, what can be inferred from Table 1 is that inconsistency indices satisfying all properties are rather rare, which means that some properties might be too restrictive.

**Fig. 2**. A timeline of advances in axiomatization of inconsistency indices (simplified). Source: authors.

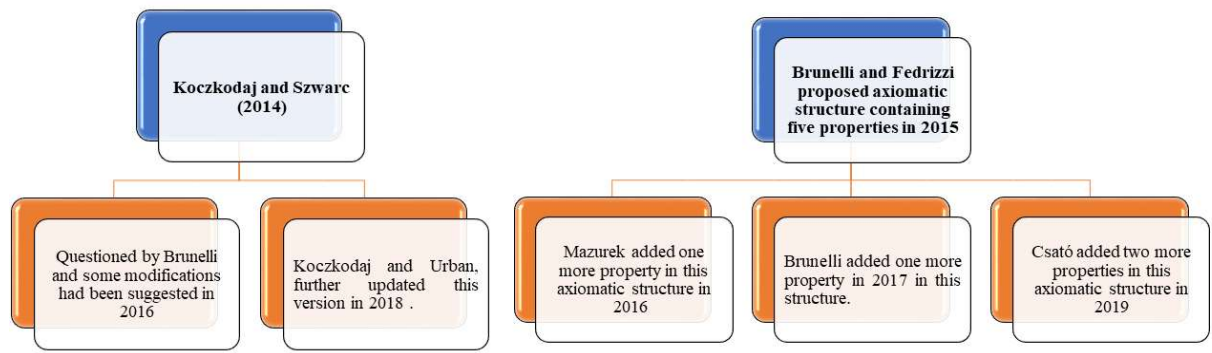

**Table 1**. Satisfaction of inconsistency indices' properties. The sign '-' means 'unknown' or 'NA'. Source: authors.

| No. | Indices | Brunelli & Fedrizzi (2015) | Mazurek (2016) | Brunelli & Fedrizzi (2017) | Koczkodaj & Urban (2018) | Csató (2018) | Csató (2019) |
|---|---|---|---|---|---|---|---|
| 1 | $CI$ Saaty (1980) | Satisfies all axioms. | Satisfies- $A_1$ to $A_5$ Dissatisfies- $A_6$ | Satisfies all axioms | Dissatisfies- $A_2$ and $A_4$ | - | - |
| 2 | $CI^*$ Peláez & Lamata 2003) | Satisfies all axioms. | Satisfies- $A_1$ to $A_5$ Dissatisfies- $A_6$ | Satisfies $A_1, A_2, A_3, A_4$ and $A_5$ | - | - | - |
| 3 | $GCI$ Aguarón & Moreno-Jiménez (2003) | Satisfies all axioms. | Satisfies- $A_1$ to $A_5$ Dissatisfies- $A_6$ | Satisfies all axioms | - | - | - |
| 4 | $KI$ Koczkodaj, (1993) | Satisfies all five axioms. | Satisfies all six axioms | Satisfies all six axioms | - | Satisfies all six axioms | Satisfies all eight axioms |
| 5 | $RE$ Barzilai (1998) | Satisfies- $A_1, A_2, A_3$ Dissatisfies- $A_4, A_5$ | Satisfies- $A_1, A_2, A_3, A_6$ Dissatisfies- $A_4, A_5$ | Satisfies- $A_1, A_2, A_3, A_6$ Dissatisfies- $A_4, A_5$ | - | - | - |
| 6 | $HCI$ Stein & Mizzi (2007) | Satisfies- $A_1, A_2, A_4$ and $A_5$ Dissatisfies- $A_3$ | - | Satisfies- $A_1, A_2, A_4, A_5$ and $A_6$ Dissatisfies- $A_3$ | - | - | - |
| 7 | $GW$ Golden & Wang (1989) | Satisfies- $A_1, A_2$ and $A_5$ Dissatisfies- $A_3$ | Satisfies- $A_1, A_2, A_5$, and $A_6$ Dissatisfies- $A_3$ | Satisfies- $A_1, A_2, A_5$ and $A_6$ Dissatisfies $A_3$ | - | - | - |
| 8 | $NI_n^\sigma$ Ramík & Korviny (2010) | Satisfies- $A_1, A_2$ and $A_5$ Dissatisfies- $A_4$ | - | Satisfies- $A_1, A_2, A_5$ and $A_6$ Dissatisfies $A_4$ | - | - | - |
| 9 | $CMSH$ Salo & Hämäläinen (1997) | Satisfies- $A_1, A_2, A_4$ and $A_5$ Dissatisfies- $A_3$ | Satisfies- $A_1, A_2, A_4$, and $A_5$ Dissatisfies- $A_3$ | Satisfies- $A_1, A_2, A_4, A_5$ and $A_6$ Dissatisfies- $A_3$ | - | - | - |
| 10 | $CI_H$ [Wu & Xu (2012)] | Satisfies all five axioms. | Satisfies- $A_1, A_2, A_3, A_4$ and $A_5$ | Satisfies- all six axioms. | - | - | - |
| 11 | $CCI$ [Kou & Lin(2014)] | Satisfies- $A_1, A_2$ and $A_5$ Dissatisfies- $A_3$ | Satisfies- $A_1, A_2, A_5$ and $A_6$ Dissatisfies- $A_3$ | Satisfies- $A_1, A_2, A_5$ and $A_6$ Dissatisfies $A_3$ | - | - | - |
| 12 | $RIC$ [Mazurek (2018)] | Satisfies- $A_1, A_2, A_4$ and $A_5$ Dissatisfies- $A_3$ | Satisfies- $A_1, A_2, A_4, A_5$ and, $A_6$ Dissatisfies- $A_3$ | Satisfies- $A_1, A_2, A_4$ and $A_5$ Dissatisfies- $A_3$ | - | - | - |
| 13 | $S$ Dixit (2018) | Satisfies all axioms. | Satisfies- $A_1 - A_5$, Dissatisfies- $A_6$ | Satisfies all axioms. | - | - | - |
| 14 | $I_{x^2}$ Fedrizzi & Ferrari (2018) | Satisfies- $A_1, A_2$, and $A_5$ | Satisfies- $A_1, A_2, A_5$ Dissatisfies- $A_6$ | Satisfies- $A_1, A_2, A_5$ and $A_6$ | - | - | - |
| 15 | $FG$ [Fedrizzi & Giove (2007)] | Satisfies all axioms. | Satisfies- $A_1 - A_5$, Dissatisfies- $A_6$ | Satisfies all axioms. | - | - | - |
| 16 | $ATI$ Grzybowski (2016) | Satisfies all axioms. | Satisfies- $A_1 - A_5$, Dissatisfies- $A_6$ | Satisfies all axioms. | - | - | - |
| 17 | $I_{CD}$ Cavallo & D'Apuzzo(2012) | Satisfies all axioms. | Satisfies- $A_1 - A_5$, Dissatisfies- $A_6$ | Satisfies all axioms. | - | - | - |
| 18 | $CI_\beta$ Sato & Tan (2023) | Satisfies- $A_1, A_2$, and $A_5$ | Satisfies- $A_1, A_2$ and $A_5$ | Satisfies- $A_1, A_2, A_5$ and $A_6$ | - | - | - |

## 5. Conclusions and future research directions

In the last decade, a number of axiomatic frameworks have been developed to examine properties of inconsistency indices in (not only) multiplicative pairwise comparisons methods. The main theoretical frameworks include those by Brunelli & Fedrizzi (2015), Csató (2018, 2019), Koczkodaj & Szwarc (2014) and Koczkodaj & Urban (2018). They share some similar features, but their main difference is that the latter three approaches are directly founded on the notion of cardinal consistency given by relation (2), which is based on triads (triples) of matrix elements. Therefore, since the cardinal consistency is defined via triads, maybe only triad-based inconsistency indices should be considered. Nevertheless, Table 1 lists the satisfaction of aforementioned properties for 18 inconsistency indices examined in the literature so far.

The aim of this paper was to provide a concise overview and discussion of the main developments in this narrow, albeit important area of operational research. Arguably, it was the study of Brunelli & Fedrizzi (2015) that sparked the debate over the axiomatization of inconsistency indices' properties. After almost 10 years, the debate is still on, and it is impossible to judge which axiomatic system (or inconsistency index) is the most suitable. For example, Csató (2019) presented an example showing that the theoretical evaluations of the axiomatic framework provided by Brunelli & Fedrizzi (2017) is probably not the best one as it doesn't work well on for just three alternatives. Therefore, even when it comes to sets of triads, the axiomatic framework offered by Brunelli & Fedrizzi (2017) is not exhaustive, meaning that they allow for very strange inconsistency indices. Unlike the prior axiomatizations, Csató (2018, 2019) had a specific objective. The first, and as far as we are aware, only works in cases when a group of axioms or characteristics have been shown to specifically establish an inconsistency index. Csató (2019) writes on page 156:

*"It is worth to note that our approach is somewhat different from the previous ones in this topic, which aimed at finding a set of suitable properties to be satisfied by any reasonable inconsistency measure. Providing an axiomatic characterization implies neither we accept all properties involved as wholly justified and unquestionable nor we reject the axioms proposed by others. Therefore, the current paper does not argue against the application of other inconsistency indices. The sole implication of our result is that if one agrees with these—rather theoretic—axioms, then the Koczkodaj inconsistency ranking remains the only choice."*

Thus, Csató (2018, 2019), in particular, verified that the set of triads has just one plausible inconsistency index. All of the earlier axiomatizations permitted the use of additional, potentially incompatible inconsistency indices. This distinction appears to be very important. All of the axiomatizations above dealt with cardinal inconsistency; however, it is possible to establish a set of rules that address both cardinal and ordinal inconsistency in this area of study.

The inconsistency indices put forth so far cannot be accepted or rejected merely because they do not adhere to a particular axiomatic structure. Each axiomatic structure has some advantages and disadvantages. Developing an error-free axiomatic structure that can be applied to PCM of any order requires a sound mathematical analysis. For example; divergent opinions exist among researchers regarding the necessity of continuity in inconsistency indices. The need for continuity of a well-chosen inconsistency index should be carefully examined because there are discontinuous functions present in literature that serve as inconsistency indices.

As for the future research, we suggest the following directions:

i) Properties of some inconsistency indices (see also Table 1) with respect to some axiomatic systems are unknown, hence filling this gap would be a valuable contribution. As can be seen, Table 1 includes 18 inconsistency indices, but Mazurek (2023) lists over 25 indices. Further on, while the properties of inconsistency indices are rather well examined with respect to axioms/properties from Brunelli & Fedrizzi (2015), satisfaction or violation of properties from Csató (2018, 2019), Koczkodaj & Szwarc (2014) or Koczkodaj & Urban (2018) is not so well established.

ii) Axioms (properties) discussed in this paper relate to pairwise comparisons in the multiplicative theoretical framework. Therefore, the set of axioms for additive or fuzzy preference relations theoretical frameworks can be postulated and discussed as well. Also, some indices might be generalized to the so called *alo-groups* (abelian linearly ordered groups), that is for all pairwise comparisons systems (multiplicative, additive and fuzzy) simultaneously, see e.g. Cavallo et al. (2023).

iii) Since there are two notions of consistency – cardinal and ordinal – it might be worthy to formulate a set of axioms for both notions simultaneously.

iv) Axioms (properties) discussed in this paper relate to inconsistency indices which measure the extent of the so called cardinal consistency given by the formula (2). However, alternative approaches to consistency such as ordinal consistency or COP (*Condition of Order Preservation*), see Bana e Costa & Vansnick (2008), or Kulakowski (2015) can be considered as well.

v) Last, but not least, axiomatic systems for incomplete or uncertain pairwise comparisons (for example in the form of interval PCs) can be investigated.

To conclude, the analysis of preferences (in the form of pairwise comparisons, or other) is heading towards the ultimate goal: making better decisions in our ever complex World.